\documentstyle[12pt]{article}
\begin{document}
\baselineskip 24pt
\begin{flushright} \ November 1996\\
SNUTP 96-122 \\
hep-th/9612028\\
\end{flushright}
\begin{center}
{\large \bf A Self-Dual Bogomol'nyi Formulation of\\ the 
 Nonlinear Schr\"odinger Equation }
\vspace{.5cm}

Phillial Oh $^{a,1}$ and Chaiho Rim $^{b,2}$   \\
$^a$ {\it Department of Physics, Sung Kyun Kwan University,
Suwon 440-746,  Korea }\\
$^b$ {\it Department of Physics, Chonbuk National University,
Chonju 561-756,  Korea }\\
\end{center}

\vspace{1.0cm}

\begin{center}
{\bf Abstract} \\
\end{center}
We obtain a self-dual formulation of the conventional nonlinear 
Schr\"odinger equation (NLSE) in the 1+1 dimension by studying 
the dimensional reduction of the self-dual Chern-Simons nonlinear 
Schr\"odinger model (NLSM) in the 2+1 dimension. It is found that 
this self-dual formulation allows us to find not only the well-known 
soliton solutions from the Bogomol'nyi bound and the Galilean boost, 
but also other soliton solutions in the presence of the background 
sources.
\\


\noindent
\hbox to 10cm{\hrulefill}
\baselineskip 12pt

\noindent $^1$ {\small
Electronic-mail: ploh@newton.skku.ac.kr}\\
$^2$ {\small
Electronic-mail: rim@phy0.chonbuk.ac.kr}

\thispagestyle{empty}
\pagebreak

\baselineskip 24pt

{\it 1. Introduction\/}:
Recently, there was a discovery of a chiral NLSE
\cite{rabe,jack}, which is related  to the Chern-Simons-matter 
theory in 2+1 dimension  through dimensional reduction and 
supports novel solitons. There appeared subsequent developments 
into the diverse directions \cite{min}, even though one of 
the original  motivations was the investigation of 
the generalized statistical behavior
\cite{hald} of many particle system in 1+1 dimension 
in relation with anyon physics
\cite{myr,wilz}. 

In this letter, we  analyze further the dimensional reduction of the 
self-dual Chern-Simons NLSM of Jackiw and Pi \cite{jack-pi},
and find a self-dual formulation of the NLSE. 
It is known in Ref. \cite{jack} that the dimensional reduction 
of the Jackiw-Pi model reproduces the the conventional NLSE 
which respects the Galilean invariance. But it was not realized 
that the form of the NLSE which emerges from this reduction is also self-dual. 
We present the derivation of this self-dual formulation here 
and study the self-dual
solitons which saturate Bogomol'nyi bound. 
This Bogomol'nyi bound reproduces 
not only the known soliton solutions but also provides a useful tool 
in finding other soliton solutions in the presence of some backgrounds.

In section 2, the self-dual formulation and its 
Bogomol'nyi soliton solutions are given. Quantum mechanical 
many-body system is considered in section 3 and 
soliton solution in the presence of backgrounds in section 4.
Section 5 is the conclusion.

{\it 2.The self-dual formulation and the  Bogomol'nyi solitons \/}:
We start from the Lagrangian which can be obtained through the 
dimensional reduction of the 2+1 dimensional self-dual Chern-Simons 
NLSM of Jackiw and Pi \cite{jack-pi}: 
\begin{equation}
{\cal L}=\psi^*(i\hbar\partial_t-\phi)\psi-\frac{\hbar^2}{2m}
\vert(\partial_x-i\xi^\prime-B)\psi\vert^2
-\frac{1}{2\nu}B^\prime (\phi+\dot\xi)\,.
\label{lag1}
\end{equation}
In the above, we denote 
$\dot\xi=\partial_t\xi$, 
$\xi^\prime =\partial_x \xi$,
the gauge fields as $A^x=\xi^\prime$, 
$A^y = B$ and $A_0 =\phi$, and assume that each field is a function of $t$
and $x$ only. We also denote the Chern-Simons
coefficient $\kappa = {1 \over 2\nu}$ and
put the coupling constant $g$ of Ref. \cite{jack-pi}
equal to ${e^2 \over 2m\kappa}$ which corresponds
to the Bogomol'nyi limit. The choice $g = -{e^2 \over 2m
\kappa}$ is equivalent to changing $\nu$ to $-\nu$.
(Charge $e$ is absorbed into the definition of gauge field).
It is to be noted that there is no imaginary number 
$i$ in front of the field $B$ unlike the gauge field
$\xi^\prime$. Hence, the matter field in this reduction has 
not only the covariant coupling with the  gauge field $\xi$ but also the 
additional non-gauge interaction  with $B$.

The above Lagrangian is invariant 
under the gauge transformation:
\begin{equation}
\psi\rightarrow e^{i\Lambda(x,t)}\psi,~
\phi\rightarrow \phi-\dot\Lambda,~
\xi\rightarrow \xi+\Lambda,~B\longrightarrow B.
\end{equation}
It also has the Galilean invariance given by
\begin{equation}
\psi\rightarrow  e^{i{m \over \hbar}(vx + {v^2  \over 2}t)} 
\psi,\quad
\xi\rightarrow \xi, ~B\rightarrow B,~
\phi\rightarrow \phi + v\xi^\prime\,,
\end{equation}
under $x\rightarrow x + v t$ and  $t\rightarrow  t$.
The parity transformation is given by
\begin{equation}
\psi\rightarrow\pm\psi,~ \phi\rightarrow\phi,~
\xi\rightarrow\xi,~B\rightarrow -B.
\end{equation}

The canonical analysis of the Lagrangian of Eq. (\ref{lag1}) 
leads to a Hamiltonian given by
\begin{equation}
H=\int dx 
\left\{\frac{\hbar^2}{2m}\vert (\partial-i\xi^\prime-B)\psi\vert^2
+\phi G\right\},
\label{hami}
\end{equation}
where the Gauss's law constraint $G$ is given by
\begin{equation}
G=\rho+\frac{1}{2\nu}B^\prime\approx 0
\end{equation}
with $\rho\equiv\psi^*\psi$. 
To achieve the self-dual formulation,
we eliminate the $\xi$ field through a phase redefinition of the
matter: $\psi\rightarrow \exp (i\xi)\psi$ 
and shift $\phi$ by $- \dot \xi$. 
Then,  solving the Gauss's law constraint, 
we obtain the Hamiltonian 
in terms of the matter field only, 
\begin{equation}
H_{eff}= \frac{\hbar^2}{2m}\int dx \left\vert
\left(\partial+\nu\int K(x-y)\rho(y) dy \right)\psi
\right\vert^2
\label{chami}
\end{equation}
where $K(x-y)=\epsilon(x-y)+c.$  
$\epsilon(x)$ is the odd-step function which is $\pm 1$ depending 
on the signature of $x$. $c$ is an arbitrary  constant.
To maintain the parity invariance, we choose $c=0$.

It can be shown \cite{jackcomm} that 
 Eq. (7) is equivalent to the usual NLSE:
 After expanding the integrand and
using the symmetry property of $\rho^3$ term, 
the equation can be rewritten as 
\begin{equation}
H_{eff}=\frac{\hbar^2}{2m}\int dx
 [\psi(x)^{*\prime}\psi(x)^\prime-2\nu
\rho^2(x)]+\frac{\hbar^2\nu^2}{6m}Q^3,
\end{equation}
where $Q=\int dx \rho(x)$. Since $Q$ is time-independent,
the $Q^3$ term contribution to the resulting Hamiltonian 
equation of motion can be absorbed into a phase redefinition
of the wave function and the usual NLSE is obtained. 

Since the Hamiltonian (\ref{chami}) is a perfect square 
(positive semi-definite), 
we can  look for  the Bogomol'nyi bound, which is
obtained by solving the first order nonlinear
differential equation
\begin{equation}
\left(\frac{d}{dx}+\nu\int_{-\infty}^\infty K(x-y)\rho(y)dy
\right)\psi(x)=0.
\label{bogo}
\end{equation}
Assuming a static solution of the form 
$\psi(x)=\sqrt{\rho(x)}$,
we have 
\begin{equation}
\frac{1}{2}(\log\rho)^{\prime}+\nu\int_{-\infty}^\infty K(x-y)\rho(y) dy=0.
\label{bogo2}
\end{equation}
Differentiating the above equation with respect to $x$, 
we find the  following 1-dimensional Liouville type equation:
\begin{equation}
\frac{1}{2}(\log\rho)^{\prime\prime}+2\nu\rho(x)=0.
\label{bogo1}
\end{equation}
The soliton solution exists for the case of $\nu>0$;
\begin{equation}
\rho(x)=\frac{a}{\cosh^2(bx)}\qquad (a>0)\,,
\label{sols}
\end{equation}
with $b^2=2a \nu$. (For negative value of $\nu$, see below).
This is the parity even solution.
We may obtain the moving soliton solution
by Galilean-boosting this,
\begin{equation}
\psi (x,t)= {\vert b\vert \over \sqrt{2 \nu}}
\exp^{i{m \over \hbar} (vx - {v^2 \over 2}t)}
{1 \over \cosh b(x -vt)} \ ,
\end{equation}
which is nothing but the well-known soliton solution 
\cite{fadd}.

The field has the normalization 
\begin{equation}
N= \int_{-\infty}^\infty |\psi|^2 dx = 
\frac{\vert b\vert}{\nu} \ .
\end{equation}
If $\vert b\vert$ is an integral multiple of $\nu$, then the 
soliton number is a positive integer. The energy of the 
moving soliton is expressed by
\begin{equation}
E_{soliton}=
i\hbar\int_{-\infty}^\infty\psi^*\dot\psi=\frac{1}{2}mv^2N,
\end{equation}
and the momentum is given by $P=mvN$.
Thus we have $E=\frac{P^2}{2m}$,
which is precisely the dispersion relation 
of the classical nonrelativistic particle.

{\it 3. Quantum mechanical consideration\/}: 
Next let us consider the system quantum mechanically:
We quantize the fields first and then reduce the phase space
using a unitary transformation \cite{rime}. 
Using the symplectic structure, 
we find;
\begin{equation}
[\psi(x,t), \psi^*(y,t)]=\delta(x-y),~~
[\xi(x,t), B^\prime(y,t)]=-2i\nu\delta(x-y).
\label{comm}
\end{equation}
The second equation gives $[\xi(x,t), B(y,t)]= i\nu K(x-y)$.
To eliminate the gauge field, we introduce a unitary transformation 
given by 
\begin{equation}
U=\exp^{\left (-i\int \rho(x)\xi(x) dx\right)}.
\label{unit}
\end{equation}
Then new  Hamiltonian $H_N = UHU^{-1}$ is given by 
\begin{equation}
H_N=\int dx \left\{\frac{\hbar^2}{2m}\left\vert
\left(\partial_x+\nu\int K(x-y)\rho(y) dy -B\right)
\psi(x)\right\vert^2
+\frac{1}{2\nu}\phi(x)B^\prime(x)\right\}.
\label{qhami}
\end{equation}
Here normal ordering is assumed.
Since $\xi$ is eliminated,  the Heisenberg equation of motion for 
$B$ yields $i \dot B=[B,H_N]=0$  and therefore,
$B$ plays the role of background field.
Without loss of generality, 
$B$ can be put to zero which respects the Galilean
symmetry. 

To examine the $N$-particle quantum mechanics of the 
Hamiltonian (\ref{qhami}), we define the $N$-body 
wave function as
\begin{equation}
\Phi(N)\equiv
<0\vert\psi(x_1)\psi(x_2)\cdots\psi(x_N)\vert N>\,.
\end{equation}
Then we arrive at the $N$-body Schr\"odinger equation:
\begin{eqnarray}
i\hbar\frac{\partial \Phi(N)}{\partial t}
&=&\frac{\hbar^2}{2m}\left\{\sum_{i=1}^{N}
\left(-\partial_i+\nu\sum_{j(\neq i)}\epsilon(x_i-x_j)\right)
\left(\partial_i+\nu\sum_{j(\neq i)}
\epsilon(x_i-x_j)\right)\right\}\Phi(N)
\nonumber\\
&=&-\frac{\hbar^2}{2m}\left\{\sum_{i=1}^{N}
\left(\partial_i^2+4\nu\sum_{j(<i)}\delta(x_i-x_j) \right) 
- \nu^2 {N (N^2-1) \over 3} \right\}
\Phi(N).
\label{manybody}
\end{eqnarray}
As expected, this Schr\"odinger equation is the 
same as the conventional one  up to a constant
which depends on the  particle number. The many-body solution 
of the above equation was obtained in \cite{lieb} and the permutation 
symmetry of the solution was considered in \cite{yang}. 

{\it 4. Soliton solution in the presence of background }:
Let us return to the classical field solution for the
case  $\nu<0$. In this case,
the field interacts with itself in the repulsive way
such that the interaction does not support 
soliton solutions. Thus, one may suspect that
an additional  attractive background source to the classical
field theory Eq. (\ref{chami}) will produce a static stable 
configuration, even though it  breaks the  Galilean invariance. 
To investigate this possibility, we add to the Hamiltonian  
$\Delta H$ given by
\begin{equation}
\Delta H=\frac{\gamma}{\nu}\int \phi(x)\delta(x) dx.
\end{equation}
This addition is equivalent to introducing 
a background contribution, $B= -\gamma \epsilon (x)$,
which provides a specific reference frame to the system 
and breaks the Galilean invariance. We can check this 
either by using  the same unitary transformation as the 
Eq. (\ref{unit}) or solving the constraint directly; 
\begin{equation}
H_N= \frac{\hbar^2}{2m} \int dx 
\left\vert
\left(\partial+\nu\int K(x-y)\rho(y) dy 
+\gamma\epsilon(x)\right)\psi
\right\vert^2\,.
\label{chami2}
\end{equation}  
The Bogomol'nyi bound is determined by  
\begin{equation}
\left(\frac{d}{dx}+\nu\int_{-\infty}^\infty K(x-y)\rho(y)dy
+\gamma\epsilon(x)\right)\psi(x)=0.
\label{bogoo1}
\end{equation}
Assuming  a solution of the form 
$\psi(x)=\sqrt{\rho(x)}$ up to a constant phase, we have 
\begin{equation}
\frac{1}{2}(\log\rho)^{\prime}+\nu\int K(x-y)\rho(y) dy
+\gamma\epsilon(x)=0.
\label{bogoo2}
\end{equation}
Differentiating the above equation, 
we find the  following Liouville type equation:
\begin{equation}
\frac{1}{2}(\log\rho)^{\prime\prime}+2\nu\rho(x)
+2\gamma\delta(x)=0.
\label{bogoo3}
\end{equation}
This equation allows the following cusp-like solution; 
\begin{equation}
\rho(x)=\frac{1}{(\alpha+\beta\vert x\vert)^2}
\label{solu}
\end{equation}
with requirements:
$\beta^2= -2\nu,~~\gamma=\frac{\beta}{\alpha}>0$.

The number density $N$ is given by
\begin{equation}
N=\int_{-\infty}^\infty\rho(x)dx=\frac{2}{\alpha\beta}.
\end{equation}
Therefore, 
we succeeded in obtaining  a static solution for $\nu<0$ and $\gamma>0$
(mutually repulsive but attractive to the background):
$\beta=\sqrt{2\vert\nu\vert}$, 
$N=\frac{\gamma}{\vert\nu\vert}$.
If $N$ is a positive integer, 
then,  $\gamma$ is an integral multiple of $\vert\nu\vert$.

However, as expected from the broken Galilean invariance, 
the moving soliton solution cannot be obtained by boosting  the above static 
solution. To cure this, we may instead introduce 
a dynamical field whose interaction respects 
the  gauge and Galilean symmetry.
This is obtained if we introduce another field 
$\chi$ and add to the Lagrangian (\ref{lag1})
\begin{equation}
\Delta{\cal L}=\chi^*(i\hbar\partial_t-\eta\phi)\chi
-\frac{\hbar^2}{2m}
\left(
\vert(\partial_x-i\eta\xi^\prime)\chi\vert^2
+g (\frac{1}{2\nu}B^\prime + \psi^* \psi) \chi^* \chi
\right)\,.
\label{lag2}
\end{equation}
$\eta$ is the charge ratio of $\chi$ to $\psi$
and $g$ is the self-coupling constant of $\chi$.
This Lagrangian can also be obtained from the $2+1$ dimensional theory.
After the elimination of the field $\xi$,  
we have the Hamiltonian as 
\begin{equation}
H_{eff}=
  \frac{\hbar^2}{2m}
\int dx \left\{
     \left\vert \left(  \partial_x +
	     \nu\int dy K(x-y) (\rho(y)+ \eta \tau(y))  
	    \right)\psi\right\vert^2
+  \vert \partial \chi\vert^2 - \eta g \tau^2 
\right\}\,,
\label{chami3}
\end{equation}
where $\tau\equiv\ \chi^* \chi$.

Let us again look for a solution which saturates the Bogomol'nyi bound:
\begin{equation}
     \left(\partial_x + \nu\int dy K(x-y) (\rho(y)+ \eta
     \tau(y))\right)\psi = 0\,.
\end{equation}
Then, the classical field  $\chi$
satisfies the NLSE,
\begin{equation}
i \hbar {\partial \chi \over \partial t} = 
( -\partial^2 - 2 \eta  g \tau )\chi\,,
\end{equation}
which allows the stationary soliton solution 
when self-coupling is attractive, $\eta g >0$; up to a trivial 
phase factor it is given as 
\begin{equation}
\chi = {|\beta| \over \sqrt{|\eta g|}}
{ e^{i \beta^2 t/\hbar} \over \cosh \beta x}\,.
\end{equation}
On this soliton background, we have the 
Bogomol'nyi bound if $\psi$ is given as 
\begin{equation}
\psi = {\sqrt{B} \over \cosh \beta x},
\end{equation}
where $B = \beta^2 ( {1 \over 2\nu} - {1 \over g}) 
>0$.
The range of  the parameters is given as 
for $\eta >0$ $(g > 0)$, $0 < \nu < {g \over 2 }$
and for $\eta <0$ $(g <0)$, $\nu < -|{g \over 2 }|$
or $\nu >0$.
The soliton possesses a total soliton number  given by
\begin{equation}
N = \int dx( \rho(x) +\eta \tau(x)) = {|\beta| \over \nu}\,.
\end{equation}

It is interesting to see that around the one soliton background,  
the other field coalesces  and forms a new soliton. 
This happens even for $\nu <-|{g \over 2}|$,
and when the masses  of the two fields are different. 
In addition,  
if we let $g \to \infty$, we reproduce the soliton solution 
for $\nu >0$ in Eq. (\ref{sols}).

{\it 5. Conclusion\/}:
 We derived the self-dual formulation of the conventional
NLSE in the 1+1
dimension by studying the dimensional reduction of the 
self-dual Chern-Simons NLSM
in the 2+1 dimension. 
The merit of this formulation is that 
the Bogomol'nyi bound and 
Galilean boost allows us to find easily the known
soliton solutions and also other soliton solutions in the 
presence of the static or soliton backgrounds.

All of these results come from the 
existence of the pair of conjugate fields 
$\xi$ and $B^\prime$ in the $1+1$ dimensional theory,
which appear in the non-gauge type of covariant derivative.
The conjugate fields disappear due to the inherent constraint but 
leave the trace of the non-local effective term.
It is to be noted that 
the Calogero system is described by 
a similar non-gauge interaction \cite{jack}, 
which emerges in the dimensional reduction scheme 
of the Chern-Simons theory 
in which  one shrinks the second spatial coordinate to vanish 
after the holomorphic 
gauge fixing is done \cite{oh}. 
In this respect, one might expect 
an intertwining relation between the non-gauge type of 
covariant derivative and 
the generalized statistics in 1+1 dimension,
which we leave for future analysis.

\begin{center}
NOTE ADDED
\end{center}

After the completion of this work, we became aware of 
Ref. \cite{andr} in which the Bogomol'nyi formulation 
of the delta-function potential problem of Eq. (\ref{manybody}) 
appeared in the investigation of the possible relation
between the Chern-Simons theory and the Calogero-Sutherland model
by using the collective-field approach of the large $N$ limit.
We thank the authors of Ref. \cite{andr} for informing us their
work.\\

\noindent
{\em Acknowledgements} \\
\indent
We would like to thank Prof. R. Jackiw for the valuable comments.
This work is supported in part 
by the Korea Science and Engineering Foundation 
through the Center for Theoretical Physics  at 
Seoul National University and the project number
(96-0702-04-01-3, 96-1400-04-01-3), 
and  by the Ministry of Education through the
Research Institute for Basic Science  (BSRI/96-1419, 96-2434).
\\

\end{document}